# Web Based Brain Volume Calculation for Magnetic Resonance Images

Kevin Karsch, Brian Grinstead, Qing He, Ye Duan

*Abstract*—Brain volume calculations are crucial in modern medical research, especially in the study of neurodevelopmental disorders. In this paper, we present an algorithm for calculating two classifications of brain volume, total brain volume (TBV) and intracranial volume (ICV). Our algorithm takes MRI data as input, performs several preprocessing and intermediate steps, and then returns each of the two calculated volumes. To simplify this process and make our algorithm publicly accessible to anyone, we have created a web-based interface that allows users to upload their own MRI data and calculate the TBV and ICV for the given data. This interface provides a simple and efficient method for calculating these two classifications of brain volume, and it also removes the need for the user to download or install any applications.

## I. INTRODUCTION

Brain volume plays an important role in the study of neuropsychiatric disorders. Brain volumes are compared between patients with disorders and controls to determine if the volume differences are related to pathogenesis [2,4]. More importantly, brain volume is used as an adjustment factor in the comparison of a specific brain structure between patients and controls. Some examples can be found in [5-7]. There are also different types of brain volumes, which have to be carefully selected when used as adjustment factors [1].

Brain volume calculation thus becomes an indispensable process in most neuroimaging analysis. Although this appears simple, to our knowledge there are no readily available applications to achieve this task. There are however several applications that aid this calculation, such as brain extraction tools, but few if any of these are web-based and most require the user to download executables and install additional libraries locally. Furthermore, the installation and use of the downloadable software is often not simple and requires the user to install many other libraries unnecessary for brain extraction and brain volume calculation.

This work was supported in part by the Thompson Center for Autism and Neurodevelopmental Research Scholar Fund, the Shumaker Biomedical Informatics Graduate Fellowship, and the University of Missouri Engineering Honors Undergraduate Research Program.

Kevin Karsch is an undergraduate student in the Department of Computer Science, University of Missouri, Columbia, MO 65201 USA (phone: 314-808-5136; e-mail: krkq35@ mizzou.edu).

Brian Grinstead is an undergraduate student in the Department of Computer Science, University of Missouri, Columbia, MO 65201 USA (phone: 314-303-8817; e-mail: bpg7vc@mizzou.edu).

Qing He is a graduate student in the Department of Computer Science, University of Missouri, Columbia, MO 65201 USA (phone: 573-814-1245; e-mail: qhgb2@ mizzou.edu).

Ye Duan is an assistant professor in the Department of Computer Science, IEEE member, University of Missouri, Columbia, MO 65201 USA (phone: 573-882-3951; e-mail: duanye@missouri.edu).

The purpose of this paper is to first develop an algorithm which will calculate two different types of brain volumes. Then, we will present a web-based tool built from this algorithm to efficiently calculate these two types of brain volumes. This web-based tool can be accessed from any ordinary web browser and is located under the Software section at http://web.missouri.edu/~duanye/Research.htm.

## II. CRITERIA AND METHODS

### A. Total Brain Volume and Intracranial Volume

We focus our volume calculations on two separate classifications of brain volume: total brain volume (TBV) and intracranial volume (ICV). While other classifications of brain volume exist, TBV and ICV are the most straightforward to calculate and are often used in brain structure shape analysis [2,4]. The definitions of TBV and ICV have some inconsistencies among medical research articles, but for our purposes we will refer to TBV as the grey and white matter of the brain excluding the ventricles [1], and ICV as the sum of grey and white brain matter including the inner and outer cerebrospinal fluid spaces [3]. With these definitions, the ICV will always be greater than or equal to TBV for any calculations done on the same MRI.

### B. MRI Image Format

MRI can be stored in numerous different formats, and there are two general classifications of formats for storing MRI. The first is a scanner format, a format in which the MRI is output from the machine that captures the images. The other is an image processing format which is obtained through a conversion of the MRI from the original scanner format. For the purposes of this paper, we are interested primarily in the image processing format of MRI, specifically the Mayo Analyze Image (Analyze 7.5) and Neuroimaging Informatics Technology Initiative (NIfTI-1.1) formats. For now, we leave the conversion from the scanner format to our supported formats up to the user. Many tools exist for this conversion, such as the open source application MRIcro, available at http://www.sph.sc.edu/comd/rorden/mricro.html.

The Analyze format consists of two files, a .hdr file and a .img file. The header contains information about the image file, such as the data type, image dimensions, and voxel scale. The NIfTI format is adapted from the Analyze format; the most significant difference between NIfTI and Analyze is that the NIfTI format includes details about any affine transformations that should be applied to the MRI before viewing [12]. Also, the NIfTI format is usually contained in a single .nii file unlike the Analyze format.

## C. Preprocessing

Initially, we allow the user to input both NIfTI and Analyze files for brain volume calculation. For our calculations, Analyze files are the most efficient and easiest to handle because they separate the pixel data from the header data with their two-file implementation, and because the data contained in the header is necessary for our calculations. More importantly, Analyze files are compatible with the tools we use during brain volume calculations. In contrast, some of the information in NIfTI files is unnecessary when calculating brain volume, and some tools required for these calculations will not work with NIfTI files. Therefore, we must first convert any NIfTI file a user uploads to the Analyze format.

We can do this accurately and efficiently using the Functional Magnetic Resonance Imaging of the Brain Software Library (FSL). Losing the extra data associated with the NIfTI file is acceptable because we are only interested in calculating the volume, and the affine transformations contained in a NIfTI file, excluding scaling, will not affect this calculation. If any scaling is applied to the NIfTI image, the scaling will be applied to the voxel scales during the conversion to account for this loss of data [8,11]. Once we have obtained the Analyze header and image pair, we can proceed to the skull stripped phase. Also, if the user inputs an Analyze file, we can skip this preprocessing stage.

## D. Brain Extraction

Before calculating the TBV and ICV for the given Analyze image, the skull must be stripped from of the image and a new Analyze file must be generated containing the newly created images that contain only the brain and not the rest of the skull and other non-brain tissue. To perform this brain extraction, we use the FSL Brain Extraction Tool (BET) because of its speed and accuracy. Information about the implementation and validation of the FSL BET can be found in [8-10]. The FSL BET takes in an Analyze image pair and outputs a new Analyze image pair with the non-brain tissue deleted (Fig. 1). We can then use this new Analyze image to begin calculating the TBV and ICV. If the user chooses to do so, he or she can extract the brain with another type of software and skip this step.

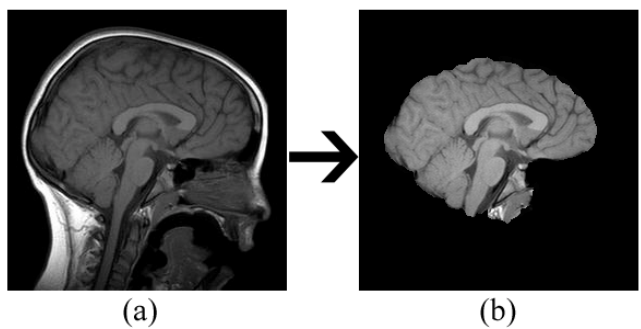

Fig. 1. Single slice of an MRI during preprocessing stages of the calculations. (a) Initial upload from user. (b) Resulting slice after brain extraction using FSL BET.

## E. Volume Calculation

To calculate the volume of the resulting skull stripped Analyze image pair, we have created an executable that takes an Analyze image pair as an argument and returns both the ICV and TBV. To do this, we first read the necessary information from the header file, such as the image dimensions, voxel scales in millimeters, and the data type (8, 16, and 32 bit integer types as well as 32 bit floating precision types are compatible with the program). Using the image dimensions, we loop through each voxel in the .img file, read the data at the given voxel and store the data into a three dimensional array. For viewing purposes, we must normalize the intensity at each voxel since we can only display 8 bits of information and some voxel values may originally contain up to 32 bits of information. To do this, we implement a contrast balancing algorithm that is used in MRIcro [13]. In short, the algorithm sets any voxel whose intensity is in and below the $2^{nd}$ percentile to 0 (black), and any voxel whose intensity is in and above the $98^{th}$ percentile to 255 (white). The remaining voxel intensities are then linearly interpolated between these two extremes to ensure that all data points lie within the 8 bit scale. Although this normalization reduces the accuracy of the original MRI intensity, the reduction is negligible and is also necessary in order to compute brain volume consistently among a variety of MRI.

With the data now normalized and stored in the array, we loop through each element in the array once more to perform several operations on the data, including the brain volume calculation. Before the loop is executed, we initialize two variables that denote the number of voxels that should be included in the TBV and ICV calculations. We also output each two dimensional image slice of the Analyze image in JPEG format to display on the website while the volume calculations are being performed. At each voxel, we check if the intensity of the voxel meets either the TBV or ICV criterion. If the criterion for either is met, we increment the TBV and ICV variables respectively. For our implementation, we consider a voxel to be part of the ICV as long as the intensity at that voxel is not zero. This means that any voxel not deleted by the skull stripping phase will be included in the ICV, which fits with our definition of ICV. Also, in our calculations, a given voxel is included in the TBV if its intensity is greater than or equal to some threshold. A threshold value of 128 works the best in many cases since it is exactly half of the range of intensities, and all grey and white brain matter is usually in this threshold. Also, since any voxel that is a member of the TBV is also a member of the ICV, we can confirm that the TBV will always be less than or equal to the ICV for each brain volume calculation. An example can be found in Fig. 2.

When the loop has finished, we obtain the number of voxels that are to be included in the TBV and ICV. Using the scaling information previously read from the header, we now multiply these totals by the scale of each voxel in the x, y and z direction. The unit associated with each scale is millimeters, so the results are always in millimeters cubed. This final calculation gives the final TBV and ICV, which we output to the user.

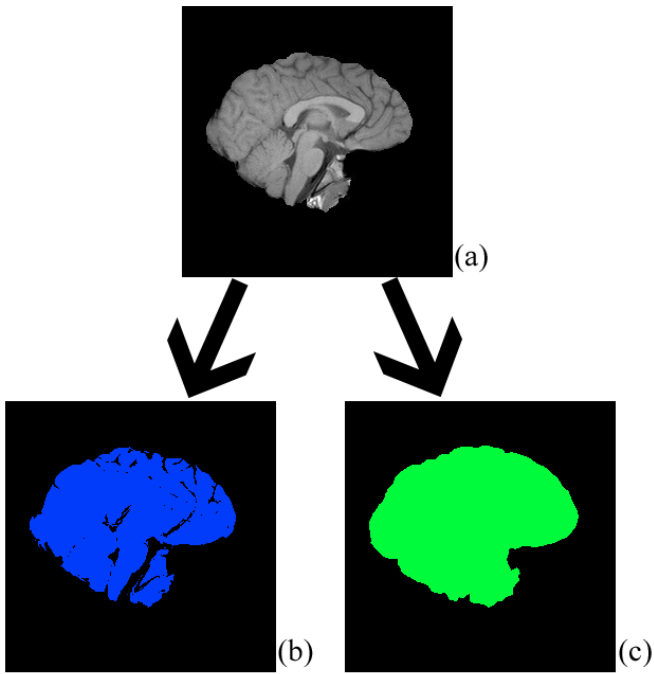

Fig. 2. An example of the differences between TBV and ICV on a single slice of an MRI. (a) Skull stripped slice obtained after preprocessing the MRI. (b) Image representation of pixels that are included in the TBV. (c) Image representation of pixels that are included in the ICV.

### III. WEB-BASED USER INTERFACE

Utilizing the algorithm provided in II, we have built a user friendly web site that allows anyone to calculate the TBV and ICV of a given MRI image set online with no configuration on their own computers. This is useful because the average user does not want to configure a library and compile source code on their computer when they will only use a couple of the utilities it provides. Our interface allows users to avoid downloading and installing these additional programs if they are interested in computing brain volume.

When a user arrives at the web page, they must either create a user name and password or login with credentials they have previously registered. Once logged in, the user will then be able to upload an MRI file (Fig. 3). After a user uploads the NIfTI or Analyze MRI file, it is stored on the web server. A program is then called to process the image set, which will perform file type conversion, brain extraction, and brain volume calculations. The TBV and ICV calculations are stored in a database so that the results can become available through the web interface the instant they are computed (Fig. 4). Also, while the volume is being calculated, the website displays the MRI slices after the brain extraction has taken place (Fig. 5). The entire process takes only a matter of minutes, and the actual execution of the all of programs embedded in the website take roughly ten seconds to three minute depending on the upload speed, size of the file, and the amount of preprocessing required.

### IV. CONCLUSION

Brain volume calculation has proven to be difficult and inefficient for some researchers, and sometimes these calculations lack accuracy as well. We have provided a simple and efficient web based method for calculating both the TBV and ICV for a given Analyze or NIfTI image, and have integrated open source applications with our own programs to attempt to calculate these values with the utmost accuracy.

We hope to also add more functionality to the website in the future to improve the ease at which users can calculate volumes. The first enhancement we will achieve is the ability to allow for additional MRI file types to be uploaded to our website, which will eliminate the need for the user to convert their MRI files to Analyze of NIfTI themselves. Another option we will give to the user is the ability to save the resulting images once we have performed the brain extraction and volume calculations so the user will be able to view a copy of the MRI showing only the TBV or ICV voxels.

Fig. 3. User interface for uploading an MRI in either Analyze of NIfTI format on the website.

Fig. 4. Results displayed on the website once an MRI has been preprocessed and the brain volume calculations have been performed.

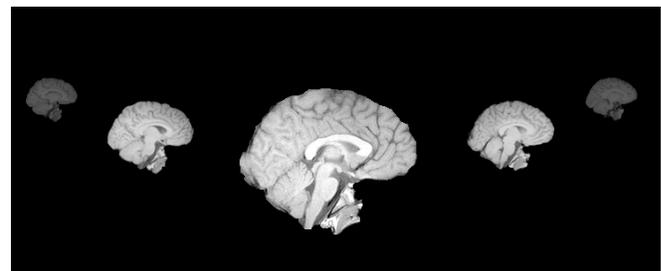

Fig. 5. Web display of several MRI slices once the brain extraction has been performed. The user is able to scroll through each slice of MRI individually.

REFERENCES

[1] L. M. O'Brien, D. A. Ziegler, C. K. Deutsch, D. N. Kennedy, J. M. Goldstein, L. J. Seidman, S. Hodge, N. Makris, V. Caviness, J. A. Frazier, and M. R. Herbert, "Adjustment for whole brain and cranial size in volumetric brain studies: a review of common ddjustment factors and statistical methods," *Harvard Review of Psychiary*, vol. 14(3), pp. 141-151, July 2006.

[2] H. C. Hazlett, M. Poe, G. Gerig, R. G. Smith, J. Provenzale, A. Ross, J. Gilmore, and J. Piven, "Magnetic resonance imaging and head circumference study of brain size in autism: birth through 2 years," *Arch. Gen. Psychiatr.*, vol. 62, pp. 1366-1376, 2005.

[3] H. Wolf, F. Kruggel, A. Hensel, L. Wahlund, T. Arendt, and H. Gertz, "The relationship between head size and intracranial volume in elderly subjects," *Brain Research*, vol. 972(1), pp. 74-80, May 2003.

[4] P. J. Anderson, D. J. Netherway, K. McGlaughlin, and D. J. David, "Intracranial volume measurement of sagittal craniosynostosis," *J. Clin. Neurosci.*, vol. 14, pp. 455-458, 2007.

[5] C. N. Vidal, R. Nicolson, T. J. DeVito, K. M. Hayashi, J. A. Geaga, D. J. Drost, P. C. Williamson, N. Rajakumar, Y. Sui, R. A. Dutton, A. W. Toga, and P. M. Thompson, "Mapping corpus callosum deficits in autism: an index of aberrant cortical connectivity," *Biological Psychiatry*, vol. 60(3), pp. 218-225, 2006.

[6] S. L. Palmer, W. E. Reddick, J. O. Glass, A. Gajjar, O. Goloubeva, and R. K. Mulhern, "Decline in corpus callosum volume among pediatric patients with medulloblastoma: longitudinal MR imaging study," *AJNR Am. J. Neuroradiol.*, vol. 23, pp. 1088–1094, 2002.

[7] R. Nicolson, T. J. Devito, C. N. Vidal, Y. Sui, K. M. Hayashi, D. J. Drost, P. C. Williamson, N. Rajakumar, A. W. Toga, and P. M. Thompson, "Detection and mapping of hippocampal abnormalities in autism," *Psychiatry Research: Neuroimaging*, vol. 148, pp. 11-21, 2006.

[8] S. M. Smith, M. Jenkinson, M. W. Woolrich, C. F. Beckmann, T. E. J. Behrens, H. Johansen-Berg, P. R. Bannister, M. De Luca, I. Drobnjak, D. E. Flitney, R. Niazy, J. Saunders, J. Vickers, Y. Zhang, N. De Stefano, J. M. Brady, and P. M. Matthews, "Advances in functional and structural MR image analysis and implementation as FSL," *NeuroImage*, vol. 23(S1), pp. 208-219, 2004.

[9] S. M. Smith, "Fast robust automated brain extraction," *Human Brain Mapping,* vol. 17(3), pp. 143-155, November 2002.

[10] S. Smith. (2008, March 4). *BET – brain extraction tool* (1st ed.) [Online]. Available: http://www.fmrib.ox.ac.uk/fsl/bet2/index.html

[11] S. Smith, and M. Jenkinson. (2008, March 4). *FSLUTILS: miscellaneous FSL image utilities* (1st ed.) [Online]. Available: http://www.fmrib.ox.ac.uk/fsl/avwutils/index.html

[12] M. Jenkinson. (2008, March 4). *NIfTI-1 Data Format* (1st ed.) [Online]. Available: http://nifti.nimh.nih.gov/nifti-1/

[13] C. Rorden, and M. Brett, "Stereotaxic display of brain lesions," *Behavioral Neurology*, vol. 12, pp. 191-200, 2000.